\documentclass[aps,prb,10pt]{revtex4-2}
\usepackage{subfig}
\usepackage{graphicx}
\usepackage{epsfig}
\usepackage{dcolumn}
\usepackage{bm}
\usepackage{times}
\usepackage{nicematrix}
\usepackage{amsmath}
\usepackage{color}

\def\beq{\begin{equation}}
\def\eeq{\end{equation}}
\def\beqn{\begin{align}}
\def\eeqn{\end{align}}

\begin{document}
\title{Lie symmetries, Jacobi last multipliers and new non-standard Lagrangians for dissipative dynamical systems}
\author{Gabriel Gonz\'alez}
\affiliation{C\'atedra CONCAYT--Universidad Aut\'onoma de San Luis Potos\'i, San Luis Potos\'i, 78000 MEXICO}
\affiliation{Coordinaci\'on para la Innovaci\'on y la Aplicaci\'on de la Ciencia y la Tecnolog\'ia, Universidad Aut\'onoma de San Luis Potos\'i,San Luis Potos\'i, 78000 MEXICO}


\begin{abstract}
 We present a new method based on Lie symmetries and Jacobi last multipliers which allows one to find many non-standard Lagrangians for dissipative dynamical systems. In particular, it is demonstrated that for every non-standard Lagrangian one can generate a new non-standard Lagrangian associated to a new equation of motion. We point out that the knowledge of Lie symmetries for a given dynamical system generates Jacobi last multipliers which can be used to obtain new non-standard Lagrangians for dissipative dynamical systems in a simple and straightforward way. We exemplify the new method by applying it to the case of the free particle and the simple harmonic oscillator in order to obtain new non-standard Lagrangians for dissipative systems.
\end{abstract}

\maketitle
\section{Introduction}
Modern physics theories ranging from classical to quantum field theory are formulated in terms of Lagrangians.\citep{havas1957range} Lagrangians play a central role in the understanding of symmetries and conservation laws of dynamical systems.\citep{mohanasubha2014interplay} In the context of classical mechanics the Lagrangian is used to describe the motion of point particles from the principle of least action. When the dynamical system is conservative, i.e. the particle moves in a region where the potential energy function is $U(x)$, the Lagrangian is obtained by simply subtracting the kinetic and potential energy of the system. For the one dimensional case we define the {\it standard} Lagrangian for a conservative system as a quadratic form with respect to $\dot{x}$,\citep{musielak2020special} which takes the following form
\begin{equation}\label{eq01}
  \mathcal{L}(t,x,\dot{x})=\frac{1}{2}\dot{x}^2-U(x)
\end{equation}
With a given Lagrangian we can obtain the equations of motion of the system by using the Euler-Lagrange equations
\begin{equation}\label{eq02}
  \frac{d}{dt}\left(\frac{\partial \mathcal{L}}{\partial\dot{x}}\right)-\frac{\partial\mathcal{L}}{\partial x}=0
\end{equation}
Substituting the Lagrangian given in equation (\ref{eq01}) into the Euler Lagrange equation we obtain Newton´s law of motion
\begin{equation}\label{eq02a}
  \ddot{x}=-\frac{dU}{dx}
\end{equation}
However, this construction is not useful for finding Lagrangians for dissipative systems.\citep{musielak2008standard,bersani2021lagrangian} The
reason is that there is not yet a consistent Lagrangian formulation for non-conservative systems.\citep{cieslinski2010direct,gonzalez2007relativistic} The problem of obtaining the Lagrangian from the equations of motion of a mechanical system is a particular case of
``The Inverse problem of the Calculus of Variations”.\citep{saunders2010thirty,douglas1941solution} In general, the equation of motion for a dissipative system is described by the following second order differential equation
\begin{equation}\label{eq02b}
  \ddot{x}=f(t,x,\dot{x})
\end{equation}
In order to obtain the Lagrangian for equation (\ref{eq02b}) we can expand the differentiation in equation (\ref{eq02}) to get\citep{gonzalez2004lagrangians}
\begin{equation}\label{eq03}
  \ddot{x}\frac{\partial^2 \mathcal{L}}{\partial\dot{x}^2}+\dot{x}\frac{\partial^2 \mathcal{L}}{\partial\dot{x}\partial x}+\frac{\partial^2 \mathcal{L}}{\partial\dot{x}\partial t}-\frac{\partial\mathcal{L}}{\partial x}=0
\end{equation}
If we differentiate equation (\ref{eq03}) with respect to $\dot{x}$ we get the following equation\citep{lopez2003constants}
\begin{equation}\label{eq04}
  \frac{\partial}{\partial\dot{x}}\left(\ddot{x}M\right)+x\frac{\partial M}{\partial x}+\frac{\partial M}{\partial t}=0
\end{equation}
where $M(t,x,\dot{x})=\partial^2 \mathcal{L}/\partial\dot{x}^2$. Equation (\ref{eq04}) is known as {\it Jacobi´s equation of the last multiplier}, where $M(t,x,\dot{x})$ is Jacobi´s last multiplier.\citep{jacobi1844sul,nucci2008jacobi,nucci2010lagrangians,carinena2015jacobi,carinena2021jacobi} Once a nontrivial solution to $M(t,x,\dot{x})$ has been found then the Lagrangian can be constructed by partially integrating $\partial^2 \mathcal{L}/\partial\dot{x}^2$, i.e.
\begin{equation}\label{eq08}
\mathcal{L}(x,\dot{x},t)=\int\int M(t,x,\dot{x})d\dot{x}d\dot{x}+Q(x,t)\dot{x}+R(x,t)
\end{equation}
where $Q(x,t)$ and $R(x,t)$ are functions that must satisfy the following condition\citep{nucci2007lagrangians}
\begin{equation}\label{eq08a}
\frac{\partial Q}{\partial t}-\frac{\partial R}{\partial x}=\int\int \frac{\partial M}{\partial x}d\dot{x}d\dot{x}-\dot{x}\int\frac{\partial M}{\partial x}d\dot{x}-\int\frac{\partial M}{\partial t}d\dot{x}-fM
\end{equation}
It is important to note that for a standard Lagrangian the Jacobi´s last multiplier does not depend explicitly on $\dot{x}$, therefore if a non-standard Lagrangian\citep{musielak2008standard} exists then the following condition most hold true, i.e.
\begin{equation}\label{eq04a}
\frac{\partial M}{\partial \dot{x}}\neq0
\end{equation}
Suppose now that we have the following equation of motion
\begin{equation}\label{eq05}
  \ddot{x}=f_0(t,x,\dot{x})-\frac{g(t,x)}{M(t,x,\dot{x})}
\end{equation}
where we assume that $M(t,x,\dot{x})\neq 0$.
If we substitute equation (\ref{eq05}) into equation (\ref{eq04}) we obtain
\begin{equation}\label{eq06}
  \frac{\partial}{\partial\dot{x}}\left(f_0M\right)+x\frac{\partial M}{\partial x}+\frac{\partial M}{\partial t}=0
\end{equation}
Equation (\ref{eq06}) tells us that if we know the non standard Lagrangian $\mathcal{L}_0$ associated with the equation of motion given by
\begin{equation}\label{eq07}
  \ddot{x}_0=f_0(t,x,\dot{x})
\end{equation}
then the non-standard Lagrangian associated with equation (\ref{eq05}) can be constructed by partially integrating $\partial^2 \mathcal{L}_0/\partial\dot{x}^2$, i.e.
\begin{equation}\label{eq08}
\mathcal{L}(t,x,\dot{x})=\int\int \frac{\partial^2 \mathcal{L}_0}{\partial\dot{x}^2}d\dot{x}d\dot{x}+Q(t,x)\dot{x}+R(t,x)
\end{equation}
{\bf Proposition} Suppose $\mathcal{L}_0$ is a non-standard Lagrangian for $\ddot{x}_0$; then there exists a non-standard Lagrangian given by
\begin{equation}\label{eqt}
\mathcal{L}(t,x,\dot{x})=\mathcal{L}_0(x,\dot{x},t)-\int g(t,x)dx
\end{equation}
which describes the following equation of motion
\begin{equation}\label{eqt1}
\ddot{x}=\ddot{x}_0-\frac{g(t,x)}{\frac{\partial^2 \mathcal{L}_0}{\partial\dot{x}^2}}
\end{equation}
{\bf Proof} By substituting the Lagrangian of equation (\ref{eqt}) into the Euler-Lagrange equation we get
\begin{equation}\label{eqt2}
 \ddot{x}\frac{\partial^2 \mathcal{L}_0}{\partial\dot{x}^2}+\dot{x}\frac{\partial^2 \mathcal{L}_0}{\partial x\partial \dot{x}}+\frac{\partial^2 \mathcal{L}_0}{\partial t\partial\dot{x}}-\frac{\partial\mathcal{L}_0}{\partial x}+g(t,x)=0
\end{equation}
and after substituting equation (\ref{eqt1}) into equation (\ref{eqt2}) and using the fact that $\mathcal{L}_0$ describes the equation of motion $\ddot{x}_0$ we get an identity which validates the proposition.\\
The method of Jacobi last multiplier was enhanced after Lie showed that his newly introduced symmetries provided a very
direct route for the calculation of Jacobi´s last multiplier, this approach, which is very easy to implement in one dimension, allows one to derive many multipliers and
therefore Lagrangians or for our case, non-standard Lagrangians.\citep{nucci2009old}  It was shown by Lie that if one knows several (at least two) Lie symmetries of the second-order
differential equation (\ref{eq07}), given by\citep{lie1893vorlesungen}
\begin{equation}\label{eql1}
\Gamma_j=V_j(t,x)\partial_t+G_j(t,x)\partial_x, \quad j=1,\cdots,r
\end{equation}
then we can derive many Jacobi last multipliers by calculating the following determinant
\begin{equation}\label{eql2}
\frac{1}{M^{(0)}_{nm}}=\det\begin{vmatrix}
                      1 & \dot{x} & f_0(t,x,\dot{x}) \\[0.5cm]
                      V_n & G_n & \frac{dG_n}{dt}-\dot{x} \frac{dV_n}{dt} \\[0.5cm]
                      V_m & G_m & \frac{dG_m}{dt}-\dot{x} \frac{dV_m}{dt}
                    \end{vmatrix}=\Delta^{(0)}_{nm}
\end{equation}
with ($n$,$m=1,\cdots,r$). Therefore, we can obtain many equations of motion $\ddot{x}_{nm}=\ddot{x}_0-g(t,x)\Delta^{(0)}_{nm}$ and their corresponding non-standard Lagrangians from the known symmetries of the original system $\ddot{x}_0$.\\
In the following sections we will present examples of the results mentioned above for the one-dimensional free particle and the simple harmonic oscillator. We will show that by knowing the symmetries and non-standard Lagrangians of a dynamical system one can find new equations of motion with their corresponding non-standard Lagrangians.
\section{Non-standard Lagrangians for the free particle}
It is very well known that the equation of motion for the free particle, i.e. $\ddot{x}=0$, has an eight-dimensional Lie symmetry algebra, $sl(3,\mathcal{R})$, generated by the following operators\citep{nucci2012lagrangian}
\begin{align*}
   & \Gamma_1=xt\partial_t+x^2\partial_x \\
  & \Gamma_2=x\partial_t \\
   & \Gamma_3=t^2\partial_t+xt\partial_x \\
   & \Gamma_4=x\partial_x \\
   & \Gamma_5=t\partial_t \\
  & \Gamma_6=\partial_t \\
   & \Gamma_7=t\partial_x \\
   & \Gamma_8=\partial_t
\end{align*}
Using equation (\ref{eql2}) we can obtain the nonzero determinants for all the possible combinations of two different operators given above. For example, let us work out the determinant $\Delta^{(0)}_{13}$ which is obtained from
\begin{equation}\label{eq09}
\Delta^{(0)}_{13}=\det\begin{vmatrix}
                      1 & \dot{x} & 0 \\[0.5cm]
                      tx & x^2 & x\dot{x}-t\dot{x}^2 \\[0.5cm]
                      t^2 & tx & x-t\dot{x}
                    \end{vmatrix}=-(t\dot{x}-x)^3
\end{equation}
It results that there are ten different determinants given by
\begin{align}\label{eq10}
   & \Delta^{(0)}_{13}=-(t\dot{x}-x)^3 \\
   & \Delta^{(0)}_{15}=-\dot{x}(t\dot{x}-x)^2 \\
   & \Delta^{(0)}_{16}=\dot{x}^2(t\dot{x}-x) \\
   & \Delta^{(0)}_{17}=-(t\dot{x}-x)^2 \\
   & \Delta^{(0)}_{18}=\dot{x}(t\dot{x}-x) \\
   & \Delta^{(0)}_{26}=-\dot{x}^3 \\
   & \Delta^{(0)}_{28}=\dot{x}^2 \\
   & \Delta^{(0)}_{38}=(t\dot{x}-x) \\
   & \Delta^{(0)}_{48}=-\dot{x} \\
   & \Delta^{(0)}_{87}=1
\end{align}
From the ten different determinants given above we see that all of them except the last one, $\Delta^{(0)}_{87}$, will give us non-standard Lagrangians, i.e. Lagrangians that are not a quadratic function of $\dot{x}$, for the following dissipative equations of motion
\begin{align}\label{eq11}
  \ddot{x}_{13}=g(x,t)(t\dot{x}-x)^3 &\quad \mathcal{L}_{13}=-\frac{1}{2t^2(t\dot{x}-x)}-\int g(x,t)dx+Q(x,t)\dot{x}+R(x,t) \\
  \ddot{x}_{15}=g(x,t)\dot{x}(t\dot{x}-x)^2 &\quad \mathcal{L}_{15}=\frac{\dot{x}}{x}\left(\ln(t\dot{x}-x)-\ln(\dot{x})\right)-\int g(x,t)dx+Q(x,t)\dot{x}+R(x,t) \\
  \ddot{x}_{16}=-g(x,t)\dot{x}^2(t\dot{x}-x) &\quad \mathcal{L}_{16}=\left(\frac{t\dot{x}}{x^2}-\frac{1}{x}\right)\left(\ln(\dot{x})-\ln(t\dot{x}-x)\right)-\int g(x,t)dx+Q(x,t)\dot{x}+R(x,t) \\
  \ddot{x}_{17}=g(x,t)(t\dot{x}-x)^2 &\quad \mathcal{L}_{17}=-\frac{1}{t^2}\ln(t\dot{x}-x)-\int g(x,t)dx+Q(x,t)\dot{x}+R(x,t) \\
  \ddot{x}_{18}=-g(x,t)\dot{x}(t\dot{x}-x) &\quad \mathcal{L}_{18}=-\frac{\dot{x}}{x}\ln(\dot{x})- \left(\frac{1}{t}-\frac{\dot{x}}{x}\right)\ln(t\dot{x}-x)+\frac{1}{t}\left(1+\ln(x)\right)-\int g(x,t)dx+Q(x,t)\dot{x}+R(x,t) \\
  \ddot{x}_{26}=g(x,t)\dot{x}^3 &\quad \mathcal{L}_{26}=\frac{-1}{2\dot{x}}-\int g(x,t)dx+Q(x,t)\dot{x}+R(x,t) \\
  \ddot{x}_{28}=-g(x,t)\dot{x}^2 &\quad \mathcal{L}_{28}=-\ln(\dot{x})-\int g(x,t)dx+Q(x,t)\dot{x}+R(x,t)  \\
  \ddot{x}_{38}=-g(x,t)(t\dot{x}-x) &\quad \mathcal{L}_{38}=\left(\frac{\dot{x}}{t}-\frac{x}{t^2}\right)\left(\ln(t\dot{x}-x)-1\right)-\int g(x,t)dx+Q(x,t)\dot{x}+R(x,t)  \\
  \ddot{x}_{48}=g(x,t)\dot{x} &\quad \mathcal{L}_{48}=\dot{x}\left(1-\ln(\dot{x})\right)-\int g(x,t)dx+Q(x,t)\dot{x}+R(x,t)
\end{align}
Note that if we make $g(x,t)=0$ we get the non-standard Lagrangians for the free particle. This result shows that for every known non-standard Lagrangian one can associate a dissipative system with variable coefficients with its own corresponding non-standard Lagrangian which reduces to the previously found case for  $g(x,t)=0$.
\section{Non-standard Lagrangians for the simple harmonic oscillator}
It is very well known that the equation of motion for the simple harmonic oscillator, i.e. $\ddot{x}=-k^2x$, has an eight-dimensional Lie symmetry algebra generated by the following operators\citep{nucci2007lagrangians}
\begin{align*}
   & \Gamma_1=\cos(kt)\partial_x-k\sin(kt)\partial_{\dot{x}} \\
  & \Gamma_2=\sin(kt)\partial_x+k\cos(kt)\partial_{\dot{x}} \\
   & \Gamma_3=x\partial_x+\dot{x}\partial_{\dot{x}} \\
   & \Gamma_4=\partial_t \\
   & \Gamma_5=\cos(2kt)\partial_t-kx\sin(2kt)\partial_x-\left(2k^2x\cos(2kt)-k\dot{x}\sin(2kt)\right)\partial_{\dot{x}} \\
  & \Gamma_6=\sin(2kt)\partial_t+kx\cos(2kt)\partial_x-\left(2k^2x\sin(2kt)+k\dot{x}\cos(2kt)\right)\partial_{\dot{x}}  \\
   & \Gamma_7=x\cos(kt)\partial_t-kx^2\sin(kt)\partial_x-\left(k^2x^2\cos(kt)+kx\dot{x}\cos(kt)+\dot{x}^2\cos(kt)\right)\partial_{\dot{x}} \\
   & \Gamma_8=x\sin(kt)\partial_t-kx^2\cos(kt)\partial_x-\left(k^2x^2\cos(kt)-kx\dot{x}\cos(kt)+\dot{x}^2\sin(kt)\right)\partial_{\dot{x}}
\end{align*}
Using equation (\ref{eql2}) we can obtain the nonzero determinants for all the possible combinations of two different operators given above. It results that there are 13 different determinants which give us non-standard Lagrangians for dissipative systems with variable coefficients. We will only list the equations of motions and their corresponding non-standard Lagrangians without the gauge functions $Q(t,x)$ and $R(x,t)$:

\begin{align}\label{12}
  & \ddot{x}_{13}=-k^2x-g(x,t)(kx\sin(kt)+\dot{x}\cos(kt))\\ & \mathcal{L}_{13}=\sec^{2}(kt)\left(\ln(kx\sin(kt)+\dot{x}\cos(kt))(kx\sin(kt)+\dot{x}\cos(kt))-\dot{x}\cos(kt)-kx\sin(kt)\right)-\int g(x,t)dx \\
  & \ddot{x}_{17}=-k^2x-g(x,t)(kx\sin(kt)+\dot{x}\cos(kt))^2\\ & \mathcal{L}_{17}=\sec^{2}(kt)\ln(kx\sin(kt)+\dot{x}\cos(kt))-\int g(x,t)dx \\
  &\ddot{x}_{18}=-k^2x-g(x,t)(kx\sin(kt)+\dot{x}\cos(kt))(kx\cos(kt)-\dot{x}\sin(kt))\\
  \begin{split}
   &\mathcal{L}_{18}=\frac{1}{kx\sin(kt)\cos(kt)}\left(\sin(kt)(kx\sin(kt)+\dot{x}\cos(kt))\ln(kx\sin(kt)+\dot{x}\cos(kt)) \right. \\
   &\left. \qquad + \cos(kt)(-\dot{x}\sin(kt)+kx)\ln(-\dot{x}\sin(kt)+kx)\right)-\int g(x,t)dx
   \end{split}\\
  &\ddot{x}_{23}=-k^2x-g(x,t)(-kx\cos(kt)+\dot{x}\sin(kt))  \\
  & \mathcal{L}_{23}=\csc^{2}(kt)\left(\ln(-kx\cos(kt)+\dot{x}\sin(kt))(-kx\cos(kt)+\dot{x}\sin(kt))-\dot{x}\sin(kt)+kx\cos(kt)\right)-\int g(x,t)dx \\
   &\ddot{x}_{28}=-k^2x-g(x,t)(-kx\cos(kt)+\dot{x}\sin(kt))^2  \\
   & \mathcal{L}_{28}=\csc^{2}(kt)\ln(kx\cos(kt)-\dot{x}\sin(kt))-\int g(x,t)dx \\
  &\ddot{x}_{34}=-k^2x-g(x,t)(\dot{x}^2+k^2x^2) \\
   & \mathcal{L}_{34}=\frac{\dot{x}}{kx}\arctan\left(\frac{\dot{x}}{kx}\right)-\frac{1}{2}\ln\left(\frac{\dot{x}^2}{k^2x^2}+1\right)-\int g(x,t)dx \\
  &\ddot{x}_{35}=-k^2x-g(x,t)\left((\dot{x}^2-k^2x^2)\cos(2kt)+2kx\dot{x}\sin(2kt)\right)\\
  \begin{split}
  & \mathcal{L}_{35}=\frac{1}{2kx\cos(2kt)}\left(2kx+(kx\sin(2kt)+\dot{x}\cos(2kt)-kx)\ln\left(\dot{x}(\sin(kt)+\cos(kt))+kx(\sin(kt)-\cos(kt))\right)-\right.\\ &\left. \qquad (\dot{x}\cos(2kt)+kx\sin(2kt)+kx)\ln\left(\dot{x}(\sin(kt)-\cos(kt))-kx(\sin(kt)+\cos(kt))\right)\right) -\int g(x,t)dx
  \end{split}\\
  &\ddot{x}_{47}=-k^2x-g(x,t)\left((\dot{x}^2+k^2x^2)(kx\sin(kt)+\dot{x}\cos(kt))\right) \\
  \begin{split}
  & \mathcal{L}_{47}=\frac{1}{2k^2x^2}(-(\dot{x}\cos(kt)+kx\sin(kt))\ln(\dot{x}^2+k^2x^2)+2(\dot{x}\sin(kt)-kx\cos(kt))\arctan\left(\frac{\dot{x}}{kx}\right)+\\
  & \qquad 2(\dot{x}\cos(kt)+kx\sin(kt))\ln(\dot{x}\cos(kt)+kx\sin(kt)))-\int g(x,t)dx
  \end{split}\\
   &\ddot{x}_{48}=-k^2x-g(x,t)\left((\dot{x}^2+k^2x^2)(-kx\cos(kt)+\dot{x}\sin(kt))\right)\\
   \begin{split}
   & \mathcal{L}_{48}=\frac{1}{2k^2x^2}(-(\dot{x}\sin(kt)-kx\cos(kt))\ln(\dot{x}^2+k^2x^2)-2(\dot{x}\cos(kt)+kx\sin(kt))\arctan\left(\frac{\dot{x}}{kx}\right)\\
   & \qquad+ 2(\dot{x}\sin(kt)-kx\cos(kt))\ln(\dot{x}\sin(kt)-kx\cos(kt)))-\int g(x,t)dx
   \end{split} \\
 &\ddot{x}_{57}=-k^2x-g(x,t)(kx\sin(kt)+\dot{x}\cos(kt))\left((\dot{x}^2-k^2x^2)\cos(2kt)+2kx\dot{x}\sin(2kt)\right) \\
 \begin{split}
 &\mathcal{L}_{57}=\frac{1}{2k^2x^2}\left(((\dot{x}(\cos(kt)-\sin(kt)))+(\cos(kt)+\sin(kt))kx)\ln((\dot{x}(\cos(kt)-\sin(kt))+(\cos(kt)+\sin(kt))kx))\right.\\
 &\left.\qquad+ ((\cos(kt)+\sin(kt))k\dot{x}-kx(\cos(kt)-\sin(kt)))\ln((\dot{x}(\cos(kt)+\sin(kt))-(\cos(kt)-\sin(kt))kx))-\right.\\
 &\left.\qquad 2(\dot{x}\cos(kt)+kx\sin(kt))\ln(\dot{x}\cos(kt)+kx\sin(kt))\right)-\int g(x,t)dx
 \end{split} \\
  &\ddot{x}_{58}=-k^2x-g(x,t)(-kx\cos(kt)+\dot{x}\sin(kt))\left((\dot{x}^2-k^2x^2)\cos(2kt)+2kx\dot{x}\sin(2kt)\right) \\
  \begin{split}
  & \mathcal{L}_{58}=\frac{1}{2k^2x^2}\left(((\dot{x}(\cos(kt)-\sin(kt)))+(\cos(kt)+\sin(kt))kx)\ln((\dot{x}(\cos(kt)-\sin(kt))+(\cos(kt)+\sin(kt))kx))-\right.\\ &\left.\qquad((\cos(kt)+\sin(kt))k\dot{x}-kx(\cos(kt)-\sin(kt)))\ln((\dot{x}(\cos(kt)+\sin(kt))-(\cos(kt)-\sin(kt))kx))+\right.\\
  &\left.\qquad 2(\dot{x}\sin(kt)-kx\cos(kt))\ln(\dot{x}\sin(kt)-kx\sin(kt))\right)-\int g(x,t)dx
  \end{split}
  \end{align}
  \begin{align}
 &\ddot{x}_{67}=-k^2x-g(x,t)(kx\sin(kt)+\dot{x}\cos(kt))^2(-kx\cos(kt)+\dot{x}\sin(kt)) \\
 &\mathcal{L}_{67}=\frac{kx\cos(kt)-\dot{x}\sin(kt)}{2k^2x^2}\left(\ln(kx\sin(kt)+\dot{x}\cos(kt))-\ln(\dot{x}\sin(kt)-kx\cos(kt))\right)-\int g(x,t)dx \\
 &\ddot{x}_{68}=-k^2x-g(x,t)(kx\sin(kt)+\dot{x}\cos(kt))(-kx\cos(kt)+\dot{x}\sin(kt))^2 \\
 & \mathcal{L}_{68}=\frac{kx\sin(kt)+\dot{x}\cos(kt)}{2k^2x^2}\left(\ln(kx\sin(kt)+\dot{x}\cos(kt))-\ln(\dot{x}\sin(kt)-kx\cos(kt))\right)-\int g(x,t)dx
\end{align}
\section{Conclusions}
In this paper we have derive non-standard Lagrangians for dissipative systems by using the Lie symmetries and the non-standard Lagrangians of non dissipative systems.
 In particular, we have demonstrated that for every non-standard Lagrangian one can generate a new non-standard Lagrangian associated to a new equation of motion. We point out that the knowledge of Lie symmetries of a given dynamical system generates Jacobi last multipliers which can be used to obtain new non-standard Lagrangians for dissipative dynamical systems in a simple and straightforward way. Our results show that the many non-standard Lagrangians that Jacobi's last multiplier may produce can be used to obtain new non-standard Lagrangians for dissipative dynamical systems with variable coefficients.

\section*{Acknowledgments}

I would like to acknowledge support by the program Cátedras Conacyt through project 1757 and from project A1-S-43579 of SEP-CONACYT Ciencia Básica and Laboratorio Nacional de Ciencia y Tecnología de Terahertz.

\end{document}